\documentclass{article}
\usepackage[preprint]{neurips_2024}

\usepackage[utf8]{inputenc} 
\usepackage[T1]{fontenc}    

\usepackage[utf8]{inputenc} 
\usepackage[T1]{fontenc}    
\usepackage{hyperref}       
\usepackage{url}            
\usepackage{booktabs}       
\usepackage{amsfonts}       
\usepackage{nicefrac}       
\usepackage{microtype}      
\usepackage{xcolor}         

\usepackage{amsmath}
\usepackage{bbding}
\usepackage{amscd,amsfonts,amsbsy,rotating}
\usepackage{balance}
\usepackage{graphicx}
\usepackage{epsfig,epstopdf}
\usepackage{subfigure}
\usepackage{multirow}
\usepackage{booktabs}
\usepackage{color,xcolor}
\usepackage{url}
\usepackage{latexsym,bm}
\usepackage{enumitem,balance,mathtools}
\usepackage{wrapfig}
\usepackage{euscript}

\usepackage[ruled,linesnumbered]{algorithm2e}
\usepackage{algorithmic}

\usepackage{ifpdf}
\usepackage{diagbox}
\usepackage{caption}
\usepackage{makecell}
\usepackage{subfigure}
\usepackage[leftcaption]{sidecap}
\usepackage{bm}
\usepackage{textcomp}
\usepackage{upgreek}

\usepackage{enumitem}
\setenumerate[1]{itemsep=0pt,partopsep=0pt,parsep=\parskip,topsep=5pt}
\setitemize[1]{itemsep=0pt,partopsep=0pt,parsep=\parskip,topsep=5pt}
\setdescription{itemsep=0pt,partopsep=0pt,parsep=\parskip,topsep=5pt}

\usepackage{inconsolata}
\usepackage{mdframed}
\usepackage{xspace}
\usepackage{amsmath}

\usepackage{wrapfig}

\newcommand{\modelname}{CodeGRAG\xspace}

\newcommand{\minisection}[1]{\vspace{0pt}\noindent\textbf{#1.}}

\title{\modelname: Bridging the Gap between Natural Language and Programming Language via Graphical Retrieval Augmented Generation}

\author{Kounianhua Du$^1$, Jizheng Chen$^1$, Renting Rui$^1$, Huacan Chai$^1$, Lingyue Fu$^1$, \\
{\bf Wei Xia$^2$, Yasheng Wang$^2$, Ruiming Tang$^2$, Yong Yu$^1$, Weinan Zhang$^1$}\\
  $^1$Shanghai Jiao Tong University, $^2$ Huawei Noah’s Ark Lab \\
  Shanghai, China\\
  \texttt{\{kounianhuadu, wnzhang\}@sjtu.edu.cn}
  }

\begin{document}

\maketitle

\begin{abstract}
  Utilizing large language models to generate codes has shown promising meaning in software development revolution. Despite the intelligence shown by the large language models, their specificity in code generation can still be improved due to the syntactic gap and mismatched vocabulary existing between natural language and programming languages. In this paper, we propose \modelname, a Graphical Retrieval Augmented Code Generation framework that bridges the gap between NL 
  and PL to enhance the performance of LLMs. \modelname builds the graphical view of code blocks based on the control flow and data flow of them to better interpret the programming domain knowledge, which can facilitate natural language based LLMs for better understanding of code syntax and serve as a bridge among different programming languages. To take the extracted structural knowledge into the foundation models, we propose 1) a hard meta-graph prompt template to transform the challenging syntax graph into informative graphical view for tuning-free models and 2) a soft prompting technique that injects the domain knowledge of programming languages into model parameters via finetuning the models with the soft signals encoded by GNN expert model. Specifically, two constraints are designed to improve the alignment and structure expressiveness, contributing to the informativeness of the single-token-sized external <GraphEmb> for enhanced code generation.
\modelname significantly improves the code generation ability of LLMs and can even offer performance gain for cross-lingual code generation. Implementation is available at https://anonymous.4open.science/r/Code-5970/ .

\end{abstract}

\section{Introduction}

In recent years, large language models (LLMs) \citep{achiam2023gpt, touvron2023llama} have shown great impact in various domains. Automated code generation emerges as a captivating frontier \citep{zheng2023codegeex, roziere2023code, shen2023pangu}, promising to revolutionize software development by enabling machines to write and optimize code with minimal human intervention.

\begin{wrapfigure}{r}{0.68\linewidth}
 	\vspace{-35pt}
 	\begin{center}
 		\includegraphics[width=\linewidth]{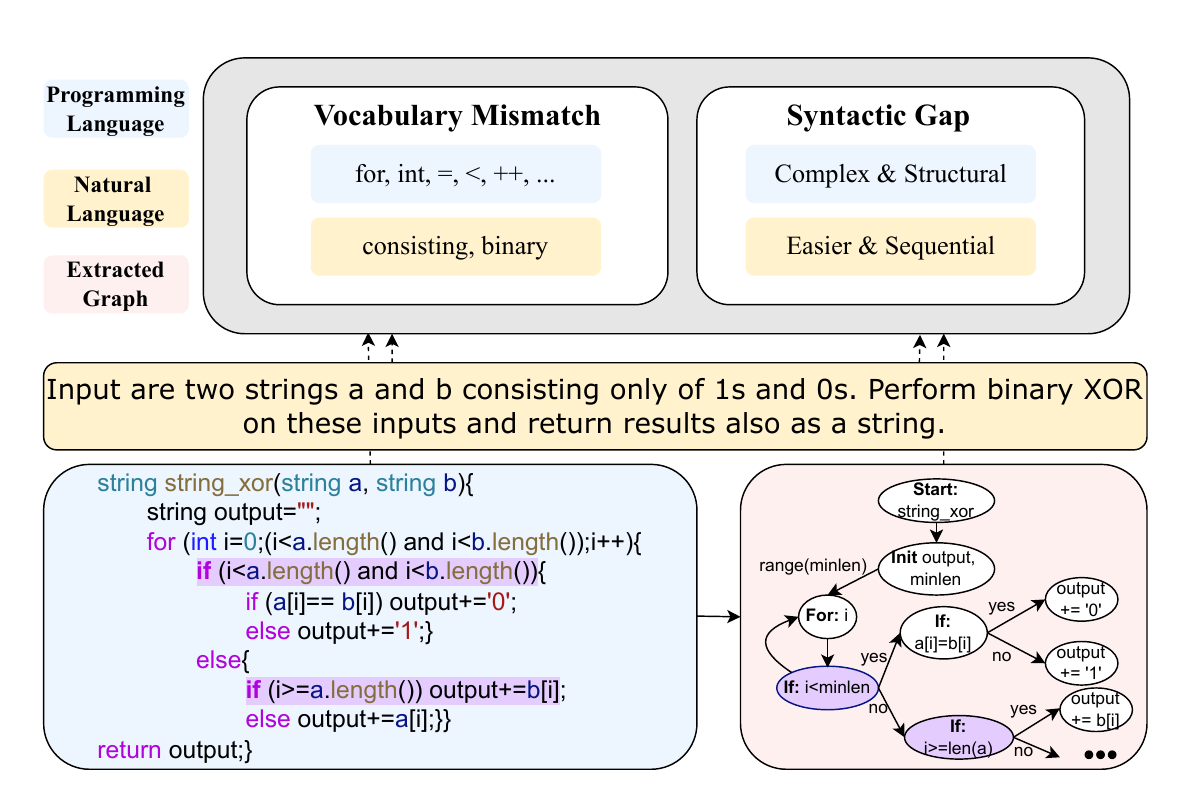}
 	\end{center}
 	\vspace{-8pt}
 	\caption{Illustration of the gap between the programming language and the natural language.}
    \vspace{-15pt}
    \label{fig:gap}
\end{wrapfigure}

Code generation involves translating natural language instructions into structured and executable code in a target programming language. However, syntactic gap and mismatched vocabulary between natural language (NL) and programming languages (PL) exist, hindering LLM's performance on code generation. As illustrated in Figure~\ref{fig:gap}, programming language (marked in blue) contains special tokens such as ``int'' or ``++'' that natural language (marked in yellow) doesn't possess with the same physical significance,
leading to vocabulary mismatch. Besides, the relations among tokens in programming languages are often structural, e.g., the complex branching and jumps, whereas natural language is arranged simply in sequential manner, leading to syntactic gap. For example, in the control flow graph of the raw code (marked in pink), two ``if'' blocks (marked in purple) are adjacent and are executed sequentially under certain condition, while appearing distant in the raw textual code.




As discussed above, the challenges of enhancing large language models for code-related tasks can be summarized into two folds. 

    

\textbf{(C1) Effective representation of programming hints to interpret the inherent logic.} 
Code, unlike natural language, possesses a well-defined structure that governs its syntax and semantics. This structure provides valuable information about the relationships between different parts of the code, the flow of execution, and the overall organization of the functions \citep{jiang2021treebert, guo2020graphcodebert}. 
LLMs regard a code block as a sequence of tokens. By ignoring the inherent structure of codes, they miss out on essential cues that could help them better understand and generate code. 
Previous RAG methods \citep{lu2022reacc, su2024evor, zhou2022docprompting} stimulate the code generation abilities of LLMs by using relevant code snippets or document as hints, failing to well represent the innate structures.

\textbf{(C2) Effective ways to digest the programming domain knowledge.} 
In addition to the well representation of the programming knowledge, the ways to inject the knowledge into the NL-based LLMs is also challenging. The structural representation of codes and the unfamiliar vocabulary could be hard to understand, which
poses a challenge to the capability of the foundation models. 

To solve the above challenges, we propose \modelname, a graphical retrieval augmented generation framework for code generation. For \textbf{(C1)}, we propose to interpret the code blocks using the graphical view composed from data-flow and control-flow of the code block to serve as the programming hints, which extracts both the semantic level and the logical level information of the code. The composed graphical view could 1) better capture the innate structural knowledge of codes for NL-based language models to understand and 2) model the innate function of code blocks that bridge different programming languages. For \textbf{(C2)}, we propose a hard meta-graph prompting technique for tuning-free models and a soft-prompting technique for tunable models. The hard meta-graph prompt summarizes the overall information of the extracted graphical view and transforms the challenging and noisy graphical representation into informative knowledge. The soft-prompting technique injects the programming domain knowledge into LLMs by tuning them with soft signals from expert GNN-encoded graphical views, under the regularization of alignment and structure-preserving objectives. After the tuning, LLMs are capable to well digest the structural graphical views by taking in the soft <GraphEmb> at single-token cost, avoiding them from the vocabulary mismatch and syntactic gap problems.

The main contributions of the paper can be summarized as follows:
\begin{itemize}[leftmargin=*]
    \item \textbf{Identification of the vocabulary mismatch and syntactic gap problems in NL2Code process.} We explain the identified problems with explicit examples and propose \modelname, a novel graphical retrieval augmented generation framework, to tackle them for enhanced code generation.
    \item \textbf{Effective graphical view to inform and stimulate the programming knowledge of LLMs.} We propose an effective graphical view 
    to serve as the programming hints, which offers more informative knowledge than the raw code block.
    \item \textbf{Effective soft prompting technique to inject the programming domain knowledge into LLMs and help them better digest the programming hints.} We propose a soft prompting technique, which injects the domain knowledge of programming languages into the model parameters via finetuning LLMs with the soft signals of GNN-encoded graphical representations. A two-hop contrastive objective for alignment along with a graph contrastive objective for structure preserving are designed to improve the expressiveness of the soft signal.
\end{itemize}

\section{Methodology}
\begin{figure*}[!t]
    \centering
    \includegraphics[width=1.0\textwidth]{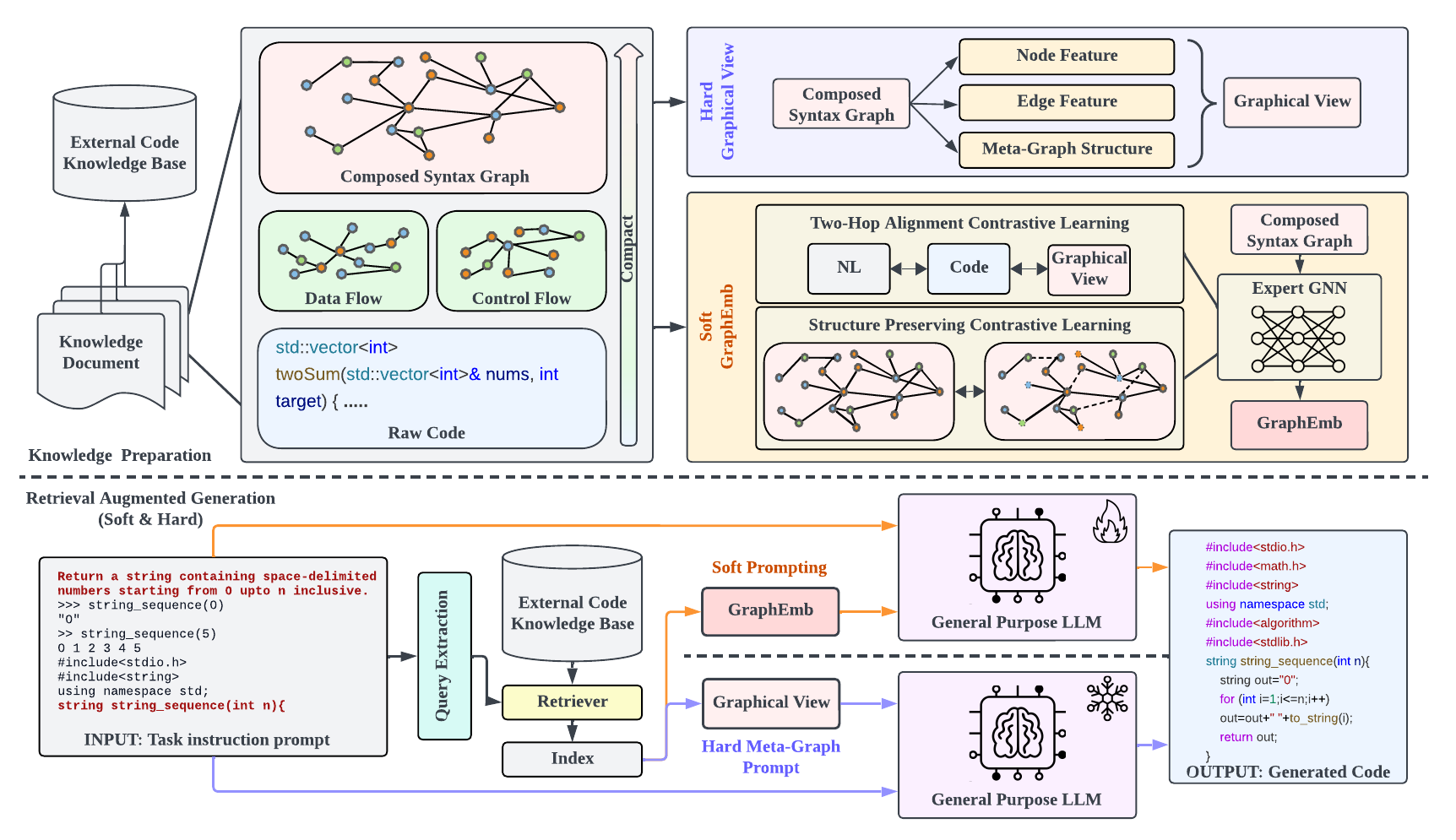}
    \vspace{-15pt}
    \caption{Overview of \modelname. \textbf{(Top) Knowledge Preparation.} We extract composed syntax graphs of external code blocks by composing the control flow and data flow of codes using the read-write signal, preserving the innate semantic and logical information. The composed graphs are then abstracted into graphical views as hard knowledge document and embedded into <GraphEmb>s as soft knowledge document. The <GraphEmb> is encoded by a pretrained GNN expert model constrained by the alignment and structure preserving objectives. \textbf{(Bottom) Retrieval Augmented Generation.} We extract query from the task input and retrieve from the external corpus. For tuning free models, we use the hard graphical view to stimulate the structural programming knowledge of LLMs for enhanced generation. For tunable models, we use the soft <GraphEmb> and inject the programming domain knowledge into LLMs parameters via finetuning them with the GNN expert signals. The expert signals informed LLMs can then produce enhanced generation.
    }
    \label{fig:illus}
    \vspace{-10pt}
\end{figure*}

\subsection{Overview}
In this paper, we leverage both generative models and retrieval models to produce results that are coherent and informed by expert graphical knowledge of the programming language.
The overall process of \modelname is illustrated in Figure~\ref{fig:illus}.

\subsection{Graphical Knowledge Base Preparation}

 To capture both the semantic and the logical information, we represent code blocks by combining the data flow graph \citep{aho2006compilers} and the control flow graph \citep{allen1970control} with the read-write signals \citep{long2022multi}, both of them are constructed on the basis of the abstract syntax tree. 

\minisection{\textbf{Abstract Syntax Tree (AST)}}  An abstract syntax tree (AST) is a tree data structure that represents the abstract syntactic structure of source code. An AST is constructed by a parser, which reads the source code and creates a tree of nodes. Each node in the tree represents a syntactic construct in the source code, such as a statement, an expression, or a declaration. 
ASTs have good compactness and can represent the structure of the source code in a clear and concise way.

\minisection{\textbf{Data Flow Graph (DFG)}}  The data flow graph (DFG) is a graphical representation of the flow of data dependencies within a program. It is a directed graph that models how data is transformed and propagated through different parts of a program.  In DFG,
nodes are operands and edges indicate data flows. 
Two types of edges are considered: 1) operation edges that
connect the nodes to be operated and the nodes that receive the operation results; 2) function edges that indicate data flows for function calls and returns. These edges connect nodes, including non-temporary operands and temporary operands, which refer to variables and constants that explicitly exist in the source code, and variables existing only in execution, respectively.

\begin{figure*}[!t]
    \centering
    \includegraphics[width=1.0\textwidth]{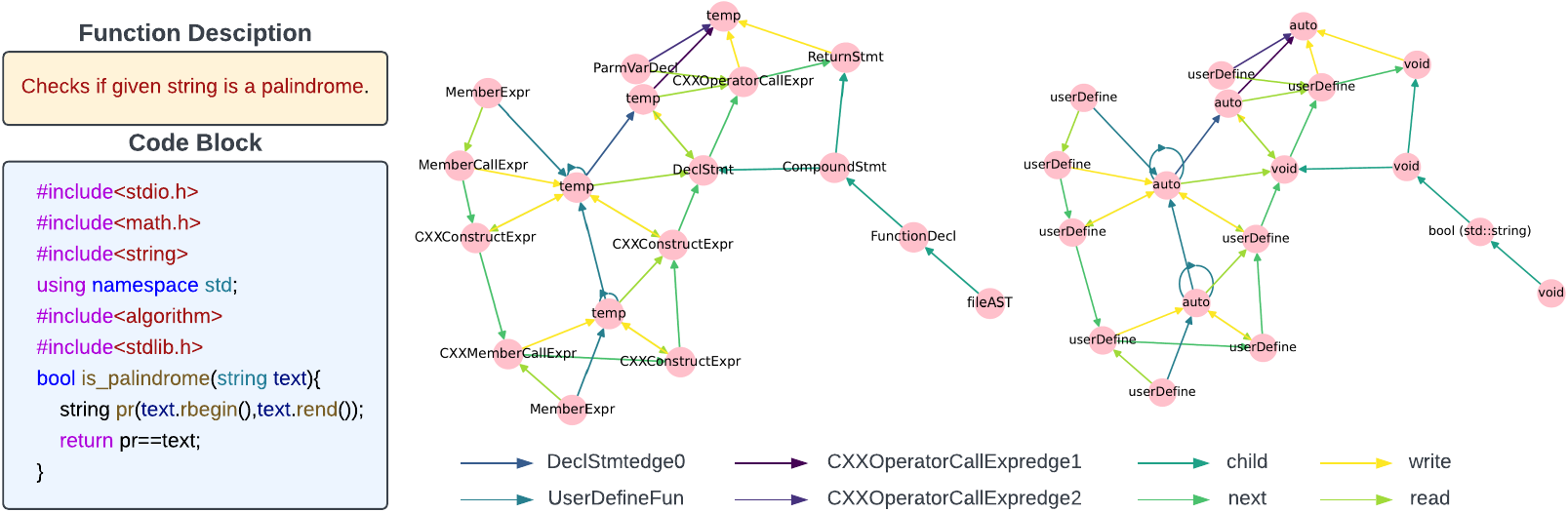}
    \vspace{-10pt}
    \caption{Illustration of the extracted composed syntax graph from the code block. The arrows in the bottom part indicate the names of different edges, which are extracted based on the ASTs.
    }
    \vspace{-10pt}
    \label{fig:graph}
\end{figure*}

\minisection{\textbf{Control Flow Graph (CFG)}} The control flow graph (CFG) is a graphical representation of the flow of control or the sequence of execution within a program. It is a directed graph that models the control relationships between different parts of a program.  Based on compiler principles,
we slightly adjust the design of CFG to better capture the key information of the program. Nodes in CFG are operations
in the source code, including standard operations, function
calls and returns. Edges indicate the execution order of operations.

\minisection{\textbf{Composed Syntax Graph}} A composed syntax graph composes the data flow graph and the control flow graph with the read-write flow existing in the code blocks. An illustration of the extracted composed syntax graph is displayed in Figure~\ref{fig:graph}. Different edge types along with their concrete names are given in colors. As for the node names,  the middle figure displays the concrete types of nodes (operands) and the right figure displays the properties of nodes.

An illustration of the composed graphical view is in Figure~\ref{fig:graph}.
After obtaining the composed syntax graphs, we use them to inform the LLMs to bridge the gap between NL and PLs, where both the semantic level and the logic level information are preserved. 

\vspace{-5pt}

\subsection{Graphical Retrieval Augmented Generation}
In this section, we discuss the ways to effectively utilize the constructed knowledge base. 


For the retrieval algorithm, we use CodeT5 to encode the query and corpus, measuring cosine distances among vectors to obtain the top-1 relevant document as the knowledge content. Since the focus of our paper is to discuss the representation of the used knowledge content but not the accuracy of ranking, we keep ranking algorithm the same for all used methods. 

For the representation of the retrieved knowledge content, since the composed syntax graph is hard to understand, we propose 1) a hard meta-graph template to transform the challenging knowledge into informative graphical view for tuning-free model and 2) a soft prompting technique to tune the foundation models for their better understanding of the graphical views with the assistance of an expert GNN model. 

\subsubsection{Hard Meta-Graph Prompt}
The original composed syntax graph of a code block could contain hundreds of nodes and edges. A full description of it could cost an overly long context, along with the understanding challenge posed by the long edge lists. Therefore, we propose to use a meta-graph\footnote{https://www.dgl.ai/dgl\_docs/generated/dgl.DGLGraph.metagraph.html} to abstract the original graph into informative graphical view. The abstracted meta-graph consists of the canonical edge types and node types, which describes the basic topology of the composed syntax graph \citep{sun2013mining}, with the textual features obtained from the ASTs contained in the node and edge features.

The template for the meta-graph is displayed as below.
\begin{mdframed}
\textit{Graph(num\_nodes=\{$node\_type: \#nodes$\},\\
\qquad \qquad num\_edges=\{$(src\_node\_type, edge\_type,$ $ dst\_node\_type): \#edges$\},\\
\qquad \qquad metagraph=[$(src\_node\_type, edge\_type,$ $ dst\_node\_type)$]})
\end{mdframed}

\begin{figure}[h]
    \centering
    \includegraphics[width=1.0\linewidth]{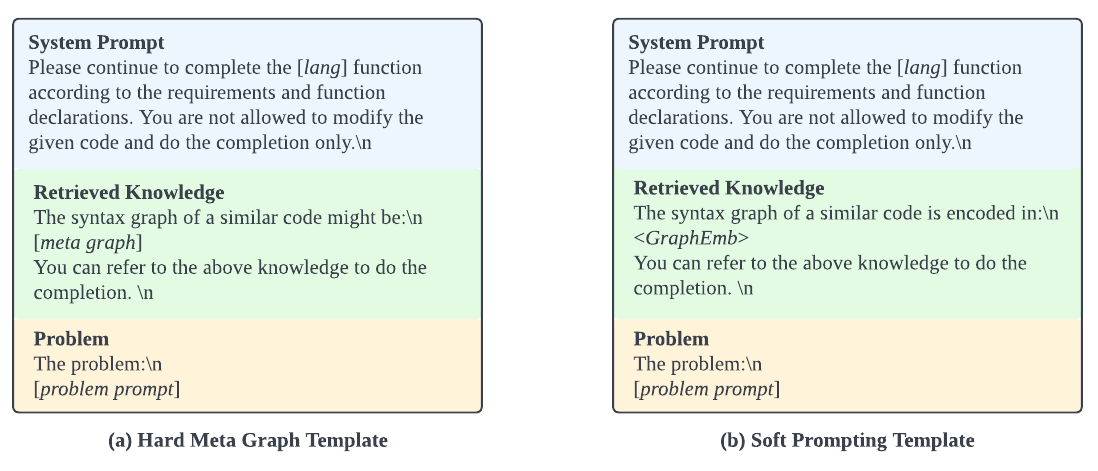}
    \vspace{-15pt}
    \caption{Prompt templates.}
    \label{fig:prompt}
\end{figure}

Then we use the meta-graph template 
to transform the retrieved graphical view into digestable knowledge and insert it into the final prompt for generation. 
As illustrated in Figure~\ref{fig:prompt} (a), the final prompt consists of three components: the system prompt illustrated in the blue part, the retrieved knowledge and hints illustrated in the green part, and the problem (including task description, function declaration, etc.) illustrated in the yellow part. The three parts are concatenated to be fed into LLMs for knowledge augmented generation.

\subsubsection{Soft Prompting with the Expert}
Directly hard prompt to the LLMs poses a challenge to the digesting capability of the backbone LLMs, suffering from the vocabulary mismatch and syntactic gap problems. To compress the graphical knowledge into model parameters and help the backbone LLMs to better understand the programming language, we propose a soft prompting technique. The overall procedure can be summarized into 
expert encoding of graphical views, finetuning with the expert signal, and inference.

We design a graph neural network to preserve the semantic and logical information of code blocks. The representation of each node $\mathbf{n}_i^{(0)}$ and edge $\mathbf{e}_i^{(0)}$ are first initialized with vectors corresponding to the node text and edge text encoded by $\phi_1$, which is a specialized coder model trained to align programming and natural languages.
A message passing process is first conducted to fuse the semantic and structural information into each node representation.
\begin{align}
    \mathbf{m}_{ij}^{(l)}  = \mathbf{W}^{(l)}&(\mathbf{n}_i^{(l-1)}\| \mathbf{e}_{ij}^{(l-1)}),\\
    Q_j^{(l)} = &\mathbf{W_Q}^{(l)}\mathbf{n}_j^{(l-1)},\\
\mathbf{K}_{ij}^{(l)} = \mathbf{W_K}^{(l)}m_{ij}^{(l)}, &\text{\quad} \mathbf{V}_{ij}^{(l)} = \mathbf{W_V}^{(l)}\mathbf{m}_{ij}^{(l)},\\
a_{ij}^{(l)}=\text{softmax}&_{i\in N(j)}\mathbf{(Q}_j^{(l)}\mathbf{K}_{ij}^{(l)}),\\
\mathbf{n}_j^{(l)} = \sum_{i\in N(j)} &a_{ij}^{(l)}\mathbf{V}_{ij}^{(l)}.
\end{align}

A global attention-based readout is then applied to obtain the graph representation:
\begin{equation}
    \mathbf{g} = \sum_{i} \text{softmax}(f_{\text{gate}}(\mathbf{n}_i^L))f_{\text{feat}}(\mathbf{n}_i^L).
\end{equation}

The expert encoding network is optimized via the contrastive learning based self-supervised training.
A two-hop contrastive learning objective is designed to align the code-NL-PL spaces, while a structure preserving contrastive learning is conducted to improve the expressiveness of the expert GNN and avoid the aligning process from being dominated by the NL side.

\begin{itemize}[leftmargin=10pt]
    \item \textbf{Two-Hop Alignment Contrastive Learning.}
There are two types of alignment to be ensured: 1) NL-Code (NC) alignment and 2) Code-Graph (CG) alignment.

We define the positive pairs and negative pairs for NC alignment purpose as:
\begin{align}
    \mathcal{I}^+_{NC} = \{\langle \mathbf{h}_i^V, \mathbf{h}_i^Q \rangle &| i\in D_{\text{train}}\}, \\
    \mathcal{I}^-_{NC} = \{\langle \mathbf{h}_i^V, \mathbf{h}_j^Q \rangle &| i\neq j, i\in D_{\text{train}}, j\in D_{\text{train}}\}.
\end{align}

And we define the positive pairs and negative pairs for CG alignment purpose as:
\begin{align}
    \mathcal{I}^+_{CG} = \{\langle \phi_1(c_i), \phi_2(g_i) \rangle | &i\in D_{\text{train}}\},\\
    \mathcal{I}^-_{CG} = \{\langle \phi_1(c_i), \phi_2(g_j) \rangle | &i\neq j, i\in D_{\text{train}}, j\in D_{\text{train}}\}.
\end{align}

\item \textbf{Structure Preserving Contrastive Learning.}
To preserve the structural information of the graphical views, we perform intra-modality contrastive learning among the graphical views and their corrupted views. Concretely, we corrupt each of the graphical view $g_i$ with the edge dropping operation to obtain its corrupted view $g_i'$. The positive pairs and negative pairs for structure-preserving purpose are then designed as:
\begin{align}
    \mathcal{I}^+_{\text{preserve}} &= \{\langle \phi_2 (g_i), \phi_2 (g_i')\rangle | i\in D_{\text{train}}\},\\
    \mathcal{I}^-_{\text{preserve}} &= \{\langle \phi_2(g_i), \phi_2(g_j') \rangle | i\neq j, i\in D_{\text{train}}, j\in D_{\text{train}}\}.
\end{align}
\end{itemize}


To help the backbone LLMs to digest the graphical views, we tune the LLMs with the expert soft signal. The prompt for finetuning consists of the system prompt, retrieved knowledge where the expert encoded graphical view is contained using a token embedding, and task prompt, as illustrated in Figure~\ref{fig:prompt} (b). After the finetuning stage, we used the tuned models to generate codes using the soft prompting template.


\section{Experiments}
\label{sec:exp}
\begin{itemize}[leftmargin=27pt]
    \item [\textbf{RQ1}] Does the proposed \modelname offer performance gain against the base model?
    \item [\textbf{RQ2}] Does the proposed graph view abstract more informative knowledge compared with the raw code block?
    \item [\textbf{RQ3}] Can soft prompting enhance the capability of the backbone LLMs? Does finetuning with the expert signals outperform the simple supervised finetuning ?
    \item [\textbf{RQ4}] Are the proposed pretraining objectives for the GNN expert effective?
    \item [\textbf{RQ5}] What is the impact of each of the components of the graphical view?
    \item [\textbf{RQ6}] How is the compatibility of the graph extraction process? 
\end{itemize}

\subsection{Setup}
\label{sec:setup}
In this paper, we evaluate \modelname with the widely used HumanEval-X\footnote{https://github.com/THUDM/CodeGeeX} \cite{zheng2023codegeex} dataset, competition-style CodeForce\footnote{https://paperswithcode.com/sota/code-generation-on-apps} and APPS\footnote{https://arxiv.org/pdf/2105.09938v3} \cite{hendrycks2021measuring} datasets. We use greedy decoding 
strategy to do the generation. The evaluation metric is Pass@1. More details of the retrieval pool and the finetuning settings can be found in Section~\ref{app:setup} in the Appendix.

\begin{table*}[t]
\centering
\caption{Results of Hard Meta-Graph Prompt on Humaneval-X. (Pass@1)}
\label{tab:res}
\resizebox{0.99\linewidth}{!}{
\begin{tabular}{c|cccc}
\hline
 Model         & Retrieved Knowledge        &  C++ &  Python \\ \hline

\multirow{5}{*}{GPT-3.5-Turbo} & N/A                        & 57.93           & 71.95              \\ \cline{2-4} 
  & Code Block \citep{nashid2023retrieval, lu2022reacc}                 & 60.37           & 72.56              \\
  & Meta-Graph                 & \textbf{62.20}           & \textbf{72.56}              \\ \cline{2-4}
  & (Multi-Lingual) Code-Block \citep{nashid2023retrieval, lu2022reacc} & 62.20           & 70.12              \\
  & (Multi-Lingual) Meta-Graph & \textbf{64.02}           & \textbf{77.44}              \\ \hline
  \multirow{5}{*}{GPT-4omini} & N/A                        & 63.41           & 78.66              \\ \cline{2-4} 
  & Code Block \citep{nashid2023retrieval, lu2022reacc}                  & 65.24           & 78.66              \\
  & Meta-Graph                 & \textbf{65.85}           & \textbf{79.88}              \\ \cline{2-4}
  & (Multi-Lingual) Code-Block \citep{nashid2023retrieval, lu2022reacc} & 65.85           & 79.27              \\
  & (Multi-Lingual) Meta-Graph & \textbf{67.07}           & \textbf{80.49}              \\ \hline
\end{tabular}
}
\end{table*}

\begin{table*}[t]
\centering
\caption{Results of Soft Prompting. (Pass@1)}
\label{tab:soft}
\begin{tabular}{c|c|cc}
\hline
Model                         & Finetune       & CodeForce (C++) & APPS (Python) \\ \hline
\multirow{3}{*}{Gemma 7b}     & N/A            & 12.83          & 5.09        \\
                              & SFT            & 14.76          & 21.09        \\
                              & Soft Prompting & 19.13          & 26.15        \\ \hline
\multirow{3}{*}{Llama2 13b}   & N/A            & 9.61          & 7.29        \\
                              & SFT            & 11.88          & 12.06        \\
                              & Soft Prompting & 13.62          & 12.74        \\ \hline
\multirow{3}{*}{CodeLlama 7b} & N/A            & 5.20          & 24.41        \\
                              & SFT            & 9.87          & 26.15        \\
                              & Soft Prompting & 11.09          & 30.26        \\ \hline
\end{tabular}
\end{table*}


\subsection{Main Results}
The main results are summarized in Table~\ref{tab:res} and Table~\ref{tab:soft}. From the results, we can draw the following conclusions.

\textbf{RQ1.} The proposed \modelname could offer performance gain against the base model, which validates the effectiveness of the proposed graphical retrieval augmented generation for code generation framework.  

\textbf{RQ2.} The model informed by the meta-graph (\modelname) could beat model informed by the raw code block. From the results, we can see that the proposed graph view could summarize the useful structural syntax information and filter out the noises, which could offer more informative knowledge hints than the raw code block. In addition, inserting the intermediate representations of codes into the prompt can stimulate the corresponding programming knowledge of LLMs.

\textbf{RQ3.} From Table~\ref{tab:soft}, we can see that finetuning with the expert soft prompting could offer more performance gain than that brought by simple supervised finetuning. This validates the effectiveness of the designed pretraining expert network and the technique of finetuning with soft prompting, which injects the programming domain knowledge into the LLMs parameters and informs the models with the structural information for gap filling.


\subsection{Impacts of the objectives for soft prompting (\textbf{RQ4})}
\begin{table*}[h]
\centering
\caption{Ablation studies on the two objectives.}
\label{tab:gnn}
\begin{tabular}{c|c|cc}
\hline
Model                         & Finetune            & CodeForce (C++) & APPS (Python) \\ \hline
\multirow{3}{*}{Gemma 7b}     & Soft Prompting      & 19.13          & 26.15        \\
                              & w/o Alignment & 7.88          & 28.58        \\
                              & w/o Structure-Preserving & 11.70          & 21.50        \\ \hline
\multirow{3}{*}{Llama2 13b}   & Soft Prompting      & 13.62          & 12.74        \\
                              & w/o Alignment & 11.79          & 10.76        \\
                              & w/o Structure-Preserving & 5.50          & 11.09        \\ \hline
\multirow{3}{*}{CodeLlama 7b} & Soft Prompting      & 11.09          & 30.26        \\
                              & w/o Alignment & 10.92          & 29.45        \\
                              & w/o Structure-Preserving & 10.66          & 26.59        \\ \hline
\end{tabular}
\end{table*}
\begin{figure}[h]
    \centering
    \includegraphics[width=1.0\linewidth]{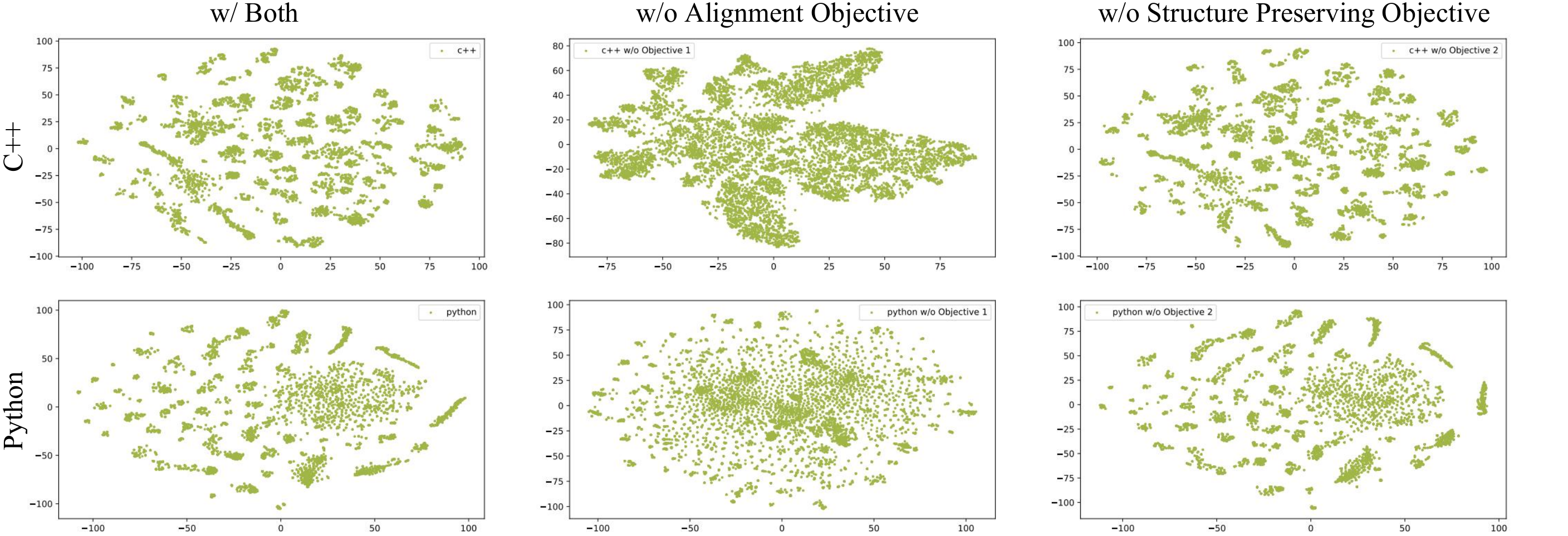}
    \caption{T-sne visualization of soft signals trained with different objectives.}
    \vspace{-10pt}
    \label{fig:tsne}
\end{figure}

To study the effectiveness of the proposed pretraining objectives for the expert GNN, we remove each objective to yield different expert GNNs and visualize the hidden representations of codes with T-SNE. The results are in Table~\ref{tab:gnn} and Figure~\ref{fig:tsne}.


From the results, we could see that both the Alignment and the Structure Preserving contribute to the expressiveness of the soft signal. The alignment pretraining objective helps to promote the alignment among natural language, programming language, and their graphical views. The structure preserving objective helps to preserve the innate data-flows and control-flows information of code blocks. The two objectives collaborate with each other to yield expressive programming domain knowledge GNN expert model, which encodes external programming knowledge and injects the knowledge into LLMs parameters.


\subsection{Impacts of the Components of the Graphical View (\textbf{RQ5})}
In this section, we adjust the inputs of the graphical components to the LLMs. Concretely, we study the information contained in node names, edge names, and the topological structure. The results are presented in Table~\ref{tab:gv}. 

\begin{wraptable}{l}{0.45\textwidth}
  \centering
  \vspace{-15pt}
    \caption{Impacts of the graph components. }
    \small
    \begin{tabular}{ccc}
    \toprule
         Datasets& Python & C++\\
         \midrule
         Edge Type Only &  73.78& 61.59\\
         Edge Type + Node Name & 75.00 & 59.76\\
         Edge Type + Node Type & 75.61 & 59.15 \\
         Edge Type + Topological & 77.44 & 64.02\\
         \bottomrule
    \end{tabular}
     \label{tab:gv}
     \vspace{-10pt}
\end{wraptable}

The edge type refers to the type of flows between operands (child, read, write, etc.), the node type refers to the type of operands (DeclStmt, temp, etc.), the node name refers to the name of the intermediate variables, and the topological information refers to the statistics of the concrete numbers of different types of edges. From the results, we can observe that 1) the edge features matter the most in constructing the structural view of code blocks for enhancement, 2) the type of nodes expresses the most in representing operands information, and 3) the overall structure of the graphical view also gives additional information.

\subsection{Compatibility Discussion (\textbf{RQ6})} 
Despite the effectiveness of the proposed graphical views to represent the code blocks, the flexibility and convenience of applying the graphical views extraction process is important for wider application of the proposed method. In this section, we discuss the compatibility of \modelname.



\begin{wraptable}{l}{0.4\textwidth}
    \centering
    \vspace{-15pt}
    \caption{Graph extraction rate (ER) of the generated results. }
    \label{tab:suc}
    \small
    \begin{tabular}{ccc}
    \toprule
         Generated Codes& Pass@1 & ER\\
         \midrule
         (C++) Code-RAG & 62.20 & 92.07\\
         (C++) \modelname &  64.02& 92.68\\
         (Python) Code-RAG & 71.95 & 91.46\\
         (Python) \modelname &  77.44& 96.95\\
         \bottomrule
    \end{tabular}
    \vspace{-10pt}
\end{wraptable}

We give the ratio of generated results that can pass the graphical views extraction process, which is denoted by Extraction Rate. 
The Pass@1 and the Extraction Rate of the generated results passing the graphical extraction process are given in Table~\ref{tab:suc}.

From the results, we could see that the extraction rates are high for codes to pass the graphical views extraction process, even under the situation where the Pass@1 ratios of the generated results are low. This indicates that the application range of the proposed method is wide. In addition, the extraction process of all the graphical views are front-end. Therefore, this extraction process applies to a wide range of code, even error code. One could also use regularization tools to reformulate the code and improve the pass rate of the extraction process.

\section{Related Work}


\minisection{\textbf{Code Search}} The code search methods can be summarized into three folds. Early methods utilize sparse search to match the query and codes \citep{hill2011improving, yang2017iecs}, which suffers from mismatched vocabulary due to the gap between NL and PL. Neural methods \citep{cambronero2019deep, gu2021multimodal} then focus on mapping the query and codes into a joint representation space for more accurate retrieval. With the success of pretrained language models, many methods propose to use pretraining tasks to improve the code understanding abilities and align different language spaces. For example, CodeBERT \citep{feng2020codebert} is pretrained on NL-PL pairs of 6 programming languages with the masked language modeling and replaced token detection task. CodeT5 \citep{wang2021codet5} supports both code-related understanding and generation tasks through bimodal dual generation. UniXcoder \citep{guo2022unixcoder} integrates the aforementioned pretraining tasks, which is a unified cross-modal pre-trained model. As retrieval augmented generation (RAG) shows its significance in promoting the quality of LLMs generation, works in code RAG start to accumulate. \cite{nashid2023retrieval, lu2022reacc} utilize the code blocks as the retrieved knowledge to inform the LLMs with similar code blocks for enhancement. \cite{zhou2022docprompting} uses the programming related document to serve as the retrieval content, injecting auxiliary external programming knowledge into the LLMs generation.

\minisection{\textbf{Code Representation}} 
Early methods regard code snippets as sequences of tokens, assuming the adjacent tokens will have strong correlations. This line of methods \citep{harer2018automated, ben2018neural, feng2020codebert, ciniselli2021empirical} take programming languages as the same with the natural language, using language models to encode the code snippets too. However, this ignoring of the inherent structure of codes leads to a loss of expressiveness. Methods that take the structural information of codes into consideration then emerge. \citet{mou2016convolutional} used convolution networks over the abstract syntax tree (AST) extracted from codes. \citet{alon2019code2vec} encoded paths sampled from the AST to represent codes. Further exploration into the graphical representation of codes \citep{allamanis2017learning} is conducted to better encode the structures of codes, where more intermediate states of the codes are considered.


\section{Limitations}
\label{sec:limit}
In this paper, we propose a graphical retrieval augmented generation method that can offer enhanced code generation. Despite the efficiency and effectiveness, there are also limitations within this work. For example, dependency on the quality of the external knowledge base could be a potential concern. The quality of the external knowledge base could be improved with regular expression extraction on the noisy texts and codes. 

\section{Conclusion}
\label{sec:conclustion}
In this paper, we locate the syntactic gap and vocabulary mismatch problems in NL2Code process and propose \modelname to alleviate the above problems for enhanced code generation.
\modelname effectively represent the programming hints to inform LLMs for code generation and improve their understanding of programming domain knowledge. Concretely, we propose to extract the semantic and logic knowledge of codes to serve as the programming hints and design proper ways to take in the knowledge: 1) a meta-graph prompt for tuning-free models and 2) a soft-prompting technique to inject the structural programming domain knowledge into the parameters of LLMs for tunable models.
Two objectives are designed to improve the alignment and structural expressiveness of the soft signals, contributing to the informativeness of the single-token-sized <GraphEmb> for enhanced code generation.
By integrating properly-represented external knowledge, \modelname enhances LLMs’ comprehension of codes and empowers them to generate code with improved accuracy. 

\begin{ack}
Use unnumbered first level headings for the acknowledgments. All acknowledgments
go at the end of the paper before the list of references. Moreover, you are required to declare
funding (financial activities supporting the submitted work) and competing interests (related financial activities outside the submitted work).
More information about this disclosure can be found at: \url{https://neurips.cc/Conferences/2024/PaperInformation/FundingDisclosure}.

Do {\bf not} include this section in the anonymized submission, only in the final paper. You can use the \texttt{ack} environment provided in the style file to automatically hide this section in the anonymized submission.
\end{ack}

\clearpage
\bibliographystyle{ACM-Reference-Format}
\bibliography{main}


\appendix

\section{Appendix}

\subsection{Implementation Details}
\label{app:setup}
For the size of the retrieval pool, we use $11,913$ C++ code snippets and $2,359$ Python code snippets. Due to the limited access, we do not use a large retrieval corpus for our experiment, which can be enlarged by other people for better performance. We also attach the graph extraction codes for both languages and all other experiment codes here: https://anonymous.4open.science/r/Code-5970/

For the finetuning details, the learning rate and weight decay for the expert GNN training are 0.001 and 1e-5, respectively. We apply 8-bit quantization and use LoRA for parameter-efficient fine-tuning. The rank of the low-rank matrices in LoRA is uniformly set to 8, alpha set to 16, and dropout is set to 0.05. The LoRA modules are uniformly applied to the Q and V parameter matrices of the attention modules in each layer of the LLM. All three models are optimized using the AdamW optimizer. For the CodeContest dataset, total $10,609$ data points are used, and for the APPS dataset, $8,691$ data samples are used to train the model.



\newpage
\section*{NeurIPS Paper Checklist}

The checklist is designed to encourage best practices for responsible machine learning research, addressing issues of reproducibility, transparency, research ethics, and societal impact. Do not remove the checklist: {\bf The papers not including the checklist will be desk rejected.} The checklist should follow the references and follow the (optional) supplemental material.  The checklist does NOT count towards the page
limit. 

Please read the checklist guidelines carefully for information on how to answer these questions. For each question in the checklist:
\begin{itemize}
    \item You should answer \answerYes{}, \answerNo{}, or \answerNA{}.
    \item \answerNA{} means either that the question is Not Applicable for that particular paper or the relevant information is Not Available.
    \item Please provide a short (1–2 sentence) justification right after your answer (even for NA). 
\end{itemize}

{\bf The checklist answers are an integral part of your paper submission.} They are visible to the reviewers, area chairs, senior area chairs, and ethics reviewers. You will be asked to also include it (after eventual revisions) with the final version of your paper, and its final version will be published with the paper.

The reviewers of your paper will be asked to use the checklist as one of the factors in their evaluation. While "\answerYes{}" is generally preferable to "\answerNo{}", it is perfectly acceptable to answer "\answerNo{}" provided a proper justification is given (e.g., "error bars are not reported because it would be too computationally expensive" or "we were unable to find the license for the dataset we used"). In general, answering "\answerNo{}" or "\answerNA{}" is not grounds for rejection. While the questions are phrased in a binary way, we acknowledge that the true answer is often more nuanced, so please just use your best judgment and write a justification to elaborate. All supporting evidence can appear either in the main paper or the supplemental material, provided in appendix. If you answer \answerYes{} to a question, in the justification please point to the section(s) where related material for the question can be found.

IMPORTANT, please:
\begin{itemize}
    \item {\bf Delete this instruction block, but keep the section heading ``NeurIPS paper checklist"},
    \item  {\bf Keep the checklist subsection headings, questions/answers and guidelines below.}
    \item {\bf Do not modify the questions and only use the provided macros for your answers}.
\end{itemize}


\begin{enumerate}

\item {\bf Claims}
    \item[] Question: Do the main claims made in the abstract and introduction accurately reflect the paper's contributions and scope?
    \item[] Answer: \answerYes{} 
    \item[] Justification: See Section~\ref{sec:exp}.
    \item[] Guidelines:
    \begin{itemize}
        \item The answer NA means that the abstract and introduction do not include the claims made in the paper.
        \item The abstract and/or introduction should clearly state the claims made, including the contributions made in the paper and important assumptions and limitations. A No or NA answer to this question will not be perceived well by the reviewers. 
        \item The claims made should match theoretical and experimental results, and reflect how much the results can be expected to generalize to other settings. 
        \item It is fine to include aspirational goals as motivation as long as it is clear that these goals are not attained by the paper. 
    \end{itemize}

\item {\bf Limitations}
    \item[] Question: Does the paper discuss the limitations of the work performed by the authors?
    \item[] Answer: \answerYes{} 
    \item[] Justification: See Section~\ref{sec:limit}.
    \item[] Guidelines:
    \begin{itemize}
        \item The answer NA means that the paper has no limitation while the answer No means that the paper has limitations, but those are not discussed in the paper. 
        \item The authors are encouraged to create a separate "Limitations" section in their paper.
        \item The paper should point out any strong assumptions and how robust the results are to violations of these assumptions (e.g., independence assumptions, noiseless settings, model well-specification, asymptotic approximations only holding locally). The authors should reflect on how these assumptions might be violated in practice and what the implications would be.
        \item The authors should reflect on the scope of the claims made, e.g., if the approach was only tested on a few datasets or with a few runs. In general, empirical results often depend on implicit assumptions, which should be articulated.
        \item The authors should reflect on the factors that influence the performance of the approach. For example, a facial recognition algorithm may perform poorly when image resolution is low or images are taken in low lighting. Or a speech-to-text system might not be used reliably to provide closed captions for online lectures because it fails to handle technical jargon.
        \item The authors should discuss the computational efficiency of the proposed algorithms and how they scale with dataset size.
        \item If applicable, the authors should discuss possible limitations of their approach to address problems of privacy and fairness.
        \item While the authors might fear that complete honesty about limitations might be used by reviewers as grounds for rejection, a worse outcome might be that reviewers discover limitations that aren't acknowledged in the paper. The authors should use their best judgment and recognize that individual actions in favor of transparency play an important role in developing norms that preserve the integrity of the community. Reviewers will be specifically instructed to not penalize honesty concerning limitations.
    \end{itemize}

\item {\bf Theory Assumptions and Proofs}
    \item[] Question: For each theoretical result, does the paper provide the full set of assumptions and a complete (and correct) proof?
    \item[] Answer: \answerNA{} 
    \item[] Justification: N/A.
    \item[] Guidelines:
    \begin{itemize}
        \item The answer NA means that the paper does not include theoretical results. 
        \item All the theorems, formulas, and proofs in the paper should be numbered and cross-referenced.
        \item All assumptions should be clearly stated or referenced in the statement of any theorems.
        \item The proofs can either appear in the main paper or the supplemental material, but if they appear in the supplemental material, the authors are encouraged to provide a short proof sketch to provide intuition. 
        \item Inversely, any informal proof provided in the core of the paper should be complemented by formal proofs provided in appendix or supplemental material.
        \item Theorems and Lemmas that the proof relies upon should be properly referenced. 
    \end{itemize}

    \item {\bf Experimental Result Reproducibility}
    \item[] Question: Does the paper fully disclose all the information needed to reproduce the main experimental results of the paper to the extent that it affects the main claims and/or conclusions of the paper (regardless of whether the code and data are provided or not)?
    \item[] Answer: \answerYes{} 
    \item[] Justification: The implementation is available at https://anonymous.4open.science/r/Code-5970/.  However, we do not provide the API needed to reproduce the results.
    \item[] Guidelines:
    \begin{itemize}
        \item The answer NA means that the paper does not include experiments.
        \item If the paper includes experiments, a No answer to this question will not be perceived well by the reviewers: Making the paper reproducible is important, regardless of whether the code and data are provided or not.
        \item If the contribution is a dataset and/or model, the authors should describe the steps taken to make their results reproducible or verifiable. 
        \item Depending on the contribution, reproducibility can be accomplished in various ways. For example, if the contribution is a novel architecture, describing the architecture fully might suffice, or if the contribution is a specific model and empirical evaluation, it may be necessary to either make it possible for others to replicate the model with the same dataset, or provide access to the model. In general. releasing code and data is often one good way to accomplish this, but reproducibility can also be provided via detailed instructions for how to replicate the results, access to a hosted model (e.g., in the case of a large language model), releasing of a model checkpoint, or other means that are appropriate to the research performed.
        \item While NeurIPS does not require releasing code, the conference does require all submissions to provide some reasonable avenue for reproducibility, which may depend on the nature of the contribution. For example
        \begin{enumerate}
            \item If the contribution is primarily a new algorithm, the paper should make it clear how to reproduce that algorithm.
            \item If the contribution is primarily a new model architecture, the paper should describe the architecture clearly and fully.
            \item If the contribution is a new model (e.g., a large language model), then there should either be a way to access this model for reproducing the results or a way to reproduce the model (e.g., with an open-source dataset or instructions for how to construct the dataset).
            \item We recognize that reproducibility may be tricky in some cases, in which case authors are welcome to describe the particular way they provide for reproducibility. In the case of closed-source models, it may be that access to the model is limited in some way (e.g., to registered users), but it should be possible for other researchers to have some path to reproducing or verifying the results.
        \end{enumerate}
    \end{itemize}

\item {\bf Open access to data and code}
    \item[] Question: Does the paper provide open access to the data and code, with sufficient instructions to faithfully reproduce the main experimental results, as described in supplemental material?
    \item[] Answer: \answerYes{} 
    \item[] Justification: The implementation is available at https://anonymous.4open.science/r/Code-5970/.
    \item[] Guidelines:
    \begin{itemize}
        \item The answer NA means that paper does not include experiments requiring code.
        \item Please see the NeurIPS code and data submission guidelines (\url{https://nips.cc/public/guides/CodeSubmissionPolicy}) for more details.
        \item While we encourage the release of code and data, we understand that this might not be possible, so “No” is an acceptable answer. Papers cannot be rejected simply for not including code, unless this is central to the contribution (e.g., for a new open-source benchmark).
        \item The instructions should contain the exact command and environment needed to run to reproduce the results. See the NeurIPS code and data submission guidelines (\url{https://nips.cc/public/guides/CodeSubmissionPolicy}) for more details.
        \item The authors should provide instructions on data access and preparation, including how to access the raw data, preprocessed data, intermediate data, and generated data, etc.
        \item The authors should provide scripts to reproduce all experimental results for the new proposed method and baselines. If only a subset of experiments are reproducible, they should state which ones are omitted from the script and why.
        \item At submission time, to preserve anonymity, the authors should release anonymized versions (if applicable).
        \item Providing as much information as possible in supplemental material (appended to the paper) is recommended, but including URLs to data and code is permitted.
    \end{itemize}

\item {\bf Experimental Setting/Details}
    \item[] Question: Does the paper specify all the training and test details (e.g., data splits, hyperparameters, how they were chosen, type of optimizer, etc.) necessary to understand the results?
    \item[] Answer: \answerYes{} 
    \item[] Justification: See Section~\ref{sec:setup} and Appendix~\ref{app:setup}.
    \item[] Guidelines:
    \begin{itemize}
        \item The answer NA means that the paper does not include experiments.
        \item The experimental setting should be presented in the core of the paper to a level of detail that is necessary to appreciate the results and make sense of them.
        \item The full details can be provided either with the code, in appendix, or as supplemental material.
    \end{itemize}

\item {\bf Experiment Statistical Significance}
    \item[] Question: Does the paper report error bars suitably and correctly defined or other appropriate information about the statistical significance of the experiments?
    \item[] Answer: \answerNo{} 
    \item[] Justification: N/A.
    \item[] Guidelines:
    \begin{itemize}
        \item The answer NA means that the paper does not include experiments.
        \item The authors should answer "Yes" if the results are accompanied by error bars, confidence intervals, or statistical significance tests, at least for the experiments that support the main claims of the paper.
        \item The factors of variability that the error bars are capturing should be clearly stated (for example, train/test split, initialization, random drawing of some parameter, or overall run with given experimental conditions).
        \item The method for calculating the error bars should be explained (closed form formula, call to a library function, bootstrap, etc.)
        \item The assumptions made should be given (e.g., Normally distributed errors).
        \item It should be clear whether the error bar is the standard deviation or the standard error of the mean.
        \item It is OK to report 1-sigma error bars, but one should state it. The authors should preferably report a 2-sigma error bar than state that they have a 96\% CI, if the hypothesis of Normality of errors is not verified.
        \item For asymmetric distributions, the authors should be careful not to show in tables or figures symmetric error bars that would yield results that are out of range (e.g. negative error rates).
        \item If error bars are reported in tables or plots, The authors should explain in the text how they were calculated and reference the corresponding figures or tables in the text.
    \end{itemize}

\item {\bf Experiments Compute Resources}
    \item[] Question: For each experiment, does the paper provide sufficient information on the computer resources (type of compute workers, memory, time of execution) needed to reproduce the experiments?
    \item[] Answer: \answerYes{} 
    \item[] Justification: See Appendix~\ref{app:setup}.
    \item[] Guidelines:
    \begin{itemize}
        \item The answer NA means that the paper does not include experiments.
        \item The paper should indicate the type of compute workers CPU or GPU, internal cluster, or cloud provider, including relevant memory and storage.
        \item The paper should provide the amount of compute required for each of the individual experimental runs as well as estimate the total compute. 
        \item The paper should disclose whether the full research project required more compute than the experiments reported in the paper (e.g., preliminary or failed experiments that didn't make it into the paper). 
    \end{itemize}
    
\item {\bf Code Of Ethics}
    \item[] Question: Does the research conducted in the paper conform, in every respect, with the NeurIPS Code of Ethics \url{https://neurips.cc/public/EthicsGuidelines}?
    \item[] Answer: \answerYes{} 
    \item[] Justification: The research conforms with the Neurips Code of Ethics.
    \item[] Guidelines:
    \begin{itemize}
        \item The answer NA means that the authors have not reviewed the NeurIPS Code of Ethics.
        \item If the authors answer No, they should explain the special circumstances that require a deviation from the Code of Ethics.
        \item The authors should make sure to preserve anonymity (e.g., if there is a special consideration due to laws or regulations in their jurisdiction).
    \end{itemize}

\item {\bf Broader Impacts}
    \item[] Question: Does the paper discuss both potential positive societal impacts and negative societal impacts of the work performed?
    \item[] Answer: \answerYes{}{} 
    \item[] Justification: See Section~\ref{sec:conclustion}.
    \item[] Guidelines:
    \begin{itemize}
        \item The answer NA means that there is no societal impact of the work performed.
        \item If the authors answer NA or No, they should explain why their work has no societal impact or why the paper does not address societal impact.
        \item Examples of negative societal impacts include potential malicious or unintended uses (e.g., disinformation, generating fake profiles, surveillance), fairness considerations (e.g., deployment of technologies that could make decisions that unfairly impact specific groups), privacy considerations, and security considerations.
        \item The conference expects that many papers will be foundational research and not tied to particular applications, let alone deployments. However, if there is a direct path to any negative applications, the authors should point it out. For example, it is legitimate to point out that an improvement in the quality of generative models could be used to generate deepfakes for disinformation. On the other hand, it is not needed to point out that a generic algorithm for optimizing neural networks could enable people to train models that generate Deepfakes faster.
        \item The authors should consider possible harms that could arise when the technology is being used as intended and functioning correctly, harms that could arise when the technology is being used as intended but gives incorrect results, and harms following from (intentional or unintentional) misuse of the technology.
        \item If there are negative societal impacts, the authors could also discuss possible mitigation strategies (e.g., gated release of models, providing defenses in addition to attacks, mechanisms for monitoring misuse, mechanisms to monitor how a system learns from feedback over time, improving the efficiency and accessibility of ML).
    \end{itemize}
    
\item {\bf Safeguards}
    \item[] Question: Does the paper describe safeguards that have been put in place for responsible release of data or models that have a high risk for misuse (e.g., pretrained language models, image generators, or scraped datasets)?
    \item[] Answer: \answerNA{} 
    \item[] Justification: N/A.
    \item[] Guidelines:
    \begin{itemize}
        \item The answer NA means that the paper poses no such risks.
        \item Released models that have a high risk for misuse or dual-use should be released with necessary safeguards to allow for controlled use of the model, for example by requiring that users adhere to usage guidelines or restrictions to access the model or implementing safety filters. 
        \item Datasets that have been scraped from the Internet could pose safety risks. The authors should describe how they avoided releasing unsafe images.
        \item We recognize that providing effective safeguards is challenging, and many papers do not require this, but we encourage authors to take this into account and make a best faith effort.
    \end{itemize}

\item {\bf Licenses for existing assets}
    \item[] Question: Are the creators or original owners of assets (e.g., code, data, models), used in the paper, properly credited and are the license and terms of use explicitly mentioned and properly respected?
    \item[] Answer: \answerYes{} 
    \item[] Justification: The existing assets used are properly cited.
    \item[] Guidelines:
    \begin{itemize}
        \item The answer NA means that the paper does not use existing assets.
        \item The authors should cite the original paper that produced the code package or dataset.
        \item The authors should state which version of the asset is used and, if possible, include a URL.
        \item The name of the license (e.g., CC-BY 4.0) should be included for each asset.
        \item For scraped data from a particular source (e.g., website), the copyright and terms of service of that source should be provided.
        \item If assets are released, the license, copyright information, and terms of use in the package should be provided. For popular datasets, \url{paperswithcode.com/datasets} has curated licenses for some datasets. Their licensing guide can help determine the license of a dataset.
        \item For existing datasets that are re-packaged, both the original license and the license of the derived asset (if it has changed) should be provided.
        \item If this information is not available online, the authors are encouraged to reach out to the asset's creators.
    \end{itemize}

\item {\bf New Assets}
    \item[] Question: Are new assets introduced in the paper well documented and is the documentation provided alongside the assets?
    \item[] Answer: \answerYes{}{} 
    \item[] Justification: Codes are public.
    \item[] Guidelines:
    \begin{itemize}
        \item The answer NA means that the paper does not release new assets.
        \item Researchers should communicate the details of the dataset/code/model as part of their submissions via structured templates. This includes details about training, license, limitations, etc. 
        \item The paper should discuss whether and how consent was obtained from people whose asset is used.
        \item At submission time, remember to anonymize your assets (if applicable). You can either create an anonymized URL or include an anonymized zip file.
    \end{itemize}

\item {\bf Crowdsourcing and Research with Human Subjects}
    \item[] Question: For crowdsourcing experiments and research with human subjects, does the paper include the full text of instructions given to participants and screenshots, if applicable, as well as details about compensation (if any)? 
    \item[] Answer: \answerNA{}{} 
    \item[] Justification: N/A.
    \item[] Guidelines:
    \begin{itemize}
        \item The answer NA means that the paper does not involve crowdsourcing nor research with human subjects.
        \item Including this information in the supplemental material is fine, but if the main contribution of the paper involves human subjects, then as much detail as possible should be included in the main paper. 
        \item According to the NeurIPS Code of Ethics, workers involved in data collection, curation, or other labor should be paid at least the minimum wage in the country of the data collector. 
    \end{itemize}

\item {\bf Institutional Review Board (IRB) Approvals or Equivalent for Research with Human Subjects}
    \item[] Question: Does the paper describe potential risks incurred by study participants, whether such risks were disclosed to the subjects, and whether Institutional Review Board (IRB) approvals (or an equivalent approval/review based on the requirements of your country or institution) were obtained?
    \item[] Answer: \answerNA{} 
    \item[] Justification: N/A.
    \item[] Guidelines:
    \begin{itemize}
        \item The answer NA means that the paper does not involve crowdsourcing nor research with human subjects.
        \item Depending on the country in which research is conducted, IRB approval (or equivalent) may be required for any human subjects research. If you obtained IRB approval, you should clearly state this in the paper. 
        \item We recognize that the procedures for this may vary significantly between institutions and locations, and we expect authors to adhere to the NeurIPS Code of Ethics and the guidelines for their institution. 
        \item For initial submissions, do not include any information that would break anonymity (if applicable), such as the institution conducting the review.
    \end{itemize}

\end{enumerate}

\end{document}